\documentclass[final,5p,times,fixfloat]{elsarticle}

\biboptions{numbers,sort&compress} %

\usepackage{graphicx}
\usepackage{amssymb} 
\usepackage{amsmath} 
\usepackage{times}
\usepackage{tabularx}

\usepackage{txfonts}
\usepackage{lineno}

\journal{Physics Letters B}

 \usepackage{ifpdf}
 \ifpdf
 \usepackage[pdftex]{hyperref}
 \else
 \usepackage[hypertex]{hyperref}
 \fi

 \hypersetup{
   pdftitle={},%
   pdfauthor={},%
   pdfsubject={},%
   pdfkeywords={},%
   pdfstartview={},%
   bookmarksopen=true, breaklinks=true, debug=true, %
   colorlinks=true, linkcolor=red, citecolor=blue, urlcolor=blue
 }

\usepackage{float} 
\usepackage[caption=false]{subfig} 

\newcount\savefnused
\newcount\savefndone

\newcommand{\savefootnote}[2][\empty]
{\ifx\empty#1\footnotemark\else\footnotemark[#1]\fi
 \global\advance\savefnused by 1
 \expandafter\xdef\csname savefnmark\the\savefnused\endcsname{\thefootnote}%
 \expandafter\xdef\csname savefntext\the\savefnused\endcsname{#2}%
}
\newcommand{\flushfootnote}{\loop\ifnum\savefndone<\savefnused
  \global\advance\savefndone by 1
  \footnotetext[\csname savefnmark\the\savefndone\endcsname]%
    {\csname savefntext\the\savefndone\endcsname}%
  \global\expandafter\let\csname savefnmark\the\savefndone\endcsname\relax
  \global\expandafter\let\csname savefntext\the\savefndone\endcsname\relax
\repeat}

\usepackage{multirow,bigdelim}

\usepackage{textcomp}
\usepackage{comment}

\usepackage{lineno}

\usepackage{float} 
\usepackage{subfig} 
\usepackage{amsmath} %
\usepackage{amstext}

\usepackage{graphicx}%
\usepackage{tabularx,ragged2e}%

\newcolumntype{Y}{>{\centering\arraybackslash}X}

\newcommand{\nn}{\nonumber}

\newcommand{\be}{\begin{equation}}
\newcommand{\ee}{\end{equation}}
\newcommand{\bea}{\begin{eqnarray}}
\newcommand{\eea}{\end{eqnarray}}
\newcommand{\balign}{\begin{align}}
\newcommand{\ealign}{\end{align}}
\newcommand{\as}{\alpha_s}

\newcommand{\st}{{\scriptscriptstyle T}}

\newcommand{\bg}{\begin{gather}}
\newcommand{\foma}{\end{gather}}

\newcommand{\noopsort}[1]{}

\def\L{\Lambda}

\def\<{\langle}
\def\>{\rangle}

\def\g{\gamma}  
  \def\D{\Delta}
\def\l{\lambda}   \def\L{\Lambda}
\def\s{\sigma}

\def\({\left(}
\def\[{\left[}
\def\){\right)}
\def\]{\right]}

\def\ln{\hbox{ln}}
\def\log{\hbox{log}}

\def\le{\left }
\def\ri{\right}

\def\gev{\rm GeV}

\newcommand{\ben}{\begin{eqnarray}}
\newcommand{\een}{\end{eqnarray}}

\newcommand{\bef}{\begin{figure}[htb]\centering}
\newcommand{\eef}{\end{figure}}

\usepackage[normalem]{ulem} 

\newcommand{\mcal}[1]{\mathcal{#1}}

\newcommand{\eq}[1]{Eq.~\eqref{#1}}

\newcommand{\dyqt}{\textsc{DYqT}}
\newcommand{\hqt}{\textsc{HqT}}

\newcommand{\icss}{\emph{iCSS}}
\newcommand{\inew}{\emph{InEW}}

\begin{document} 
\begin{frontmatter}

\title{Matching factorization theorems with an inverse-error weighting}
\author[a]{Miguel G. Echevarria}
\author[b]{Tomas Kasemets}
\author[c]{Jean-Philippe Lansberg}
\author[d]{Cristian Pisano}
\author[e]{Andrea Signori}

\address[a]{INFN, Sezione di Pavia, Via Bassi 6, 27100 Pavia, Italy}
\address[b]{PRISMA Cluster of Excellence \& Mainz Institute for Theoretical Physics\\ Johannes Gutenberg University, 55099 Mainz, Germany}
\address[c]{IPNO, CNRS-IN2P3, Univ. Paris-Sud, Universit\'e Paris-Saclay, 91406 Orsay Cedex, France}
\address[d]{Dipartimento di Fisica, Universit\`a di Cagliari and INFN,
Sezione di Cagliari, \\ Cittadella Universitaria, I-09042 Monserrato (CA), Italy}
\address[e]{Theory Center, Thomas Jefferson National Accelerator Facility \\ 12000 Jefferson Avenue, Newport News, VA 23606, USA}

\begin{abstract}

{\small
We propose a new fast method to match factorization theorems applicable in different kinematical regions, such as the transverse-momentum-dependent and the collinear factorization theorems in Quantum Chromodynamics.
At variance with well-known approaches relying on their simple addition and subsequent subtraction of double-counted contributions, ours simply builds on their weighting using the theory uncertainties deduced from the factorization theorems themselves.
This allows us to estimate the unknown complete matched cross section from an inverse-error-weighted average. 
The method is simple and provides an evaluation of the theoretical uncertainty of the matched cross section associated with the uncertainties from the power corrections to the factorization theorems
(additional uncertainties, such as the nonperturbative ones, should be added for a proper comparison with experimental data).
Its usage is illustrated with several basic examples, such as $Z$ boson, $W$ boson, $H^0$ boson and Drell-Yan lepton-pair production in hadronic collisions, and compared to the state-of-the-art Collins-Soper-Sterman subtraction scheme.
It is also not limited to the transverse-momentum spectrum, and can straightforwardly be extended to match any (un)polarized cross section differential in other variables, including multi-differential measurements.
}
\end{abstract}

%
\end{frontmatter}
%

\section{Motivation}
\label{s:motivation}

In processes with a hard scale $Q$ and a measured transverse momentum $q_\st$, for instance the mass and the transverse momentum of an electroweak boson produced in proton-proton collisions, the $q_\st$-differential cross section can be expressed through two different factorization theorems.
For small $q_\st \ll Q$, the transverse-momentum-dependent (TMD) factorization applies and the cross section is factorized in terms of TMD parton distribution/fragmentation functions  
(TMDs thereafter)~\cite{Collins:2011zzd,GarciaEchevarria:2011rb,Echevarria:2012js}. 
The evolution of the TMDs resums the large logarithms of $Q/q_\st$~\cite{Aybat:2011zv,Echevarria:2012pw,Echevarria:2014rua}. 
For large $q_\st \sim Q \gg m$, with $m$ a hadronic mass of the order of 1~GeV, there is only one hard scale in the process and the collinear factorization is the appropriate framework. 
The cross section is then written in terms of (collinear) parton distribution/fragmentation functions (PDFs/FFs). 
In order to describe the full $q_\st$ spectrum, the TMD and collinear factorization theorems must properly be matched in the intermediate region.

Many recent works on TMD phenomenology and extractions of TMDs from data did not take into account the matching with fixed-order collinear calculations for increasing transverse momentum (see e.g. Refs.~\cite{Scimemi:2017etj,Bacchetta:2017gcc}).
Such a matching is one of the compelling milestones for the next generation of TMD analyses and more generally for a thorough understanding of TMD observables~\cite{Angeles-Martinez:2015sea}.
In addition, it has recently been shown that the precisely measured transverse-momentum spectrum of $Z$ boson at the LHC does not completely agree with collinear-based NNLO computations\footnote{See \href{https://gsalam.web.cern.ch/gsalam/talks/repo/2016-03-SB+SLAC-SLAC-precision.pdf}{https://gsalam.web.cern.ch/gsalam/talks/repo/2016-03-SB+SLAC-SLAC-precision.pdf}}, hinting at possible higher-twist contributions at the per-cent level.
Thus having a reliable estimation of the matching uncertainty from power corrections is very opportune.

This work contributes to this effort by introducing a new approach, whose main features are its simplicity and its easy and fast implementation in phenomenological analyses (fits and/or Monte Carlo event generators). 
In addition, this scheme provides an automatic estimate of the theoretical uncertainty associated to the matching procedure. 
All these are crucial features in light of the computational demands of global TMD analyses and event generation for the next generation of experiments~\cite{Dudek:2012vr,Accardi:2012qut,Brodsky:2012vg,Blondel:2011fua}. 
 
As we will show, it yields compatible results with other mainstream approaches in the literature, such as the improved Collins-Soper-Sterman (CSS) scheme~\cite{Collins:2016hqq} (see also Ref.~\cite{Gamberg:2017jha}), which refines the original CSS subtraction approach~\cite{Collins:1981uk,Collins:1981uw,Collins:1984kg,Arnold:1990yk}. 
The latter, in simple terms, is based on adding the TMD-based resummed ($\cal W$) and collinear-based fixed-order ($\cal Z$) results, and then subtracting the double-counted contributions ($\cal A$).
The improved CSS (\icss) approach enforces the necessary cancellations for the subtraction method to work.

Other methods have been introduced in the framework of soft-collinear effective theory by using profile functions for the resummation scales in order to obtain analogous cancellations to those in the \icss method, see e.g. Refs.~\cite{Ligeti:2008ac,Abbate:2010xh,Stewart:2013faa,Neill:2015roa}. 
One can also find other schemes to match TMD and collinear frameworks, e.g. Refs.~\cite{Berger:2004cc,Nadolsky:2002jr,Nadolsky:1999kb}.

In the scheme we introduce, no cancellation between the TMD-based resummed contribution, $\cal W$, and the collinear-based fixed-order contribution, $\cal Z$, is needed. 
We simply avoid the double counting (and therewith the subtraction of $\cal A$) by weighting both contributions to the matched cross section, with the condition that the weights add up to unity.
This renders the computation of the matched cross section very easy to implement. 
Clearly, the weights cannot be arbitrary and should ensure that, in their respective domains of applicability, the predictions of both factorization theorems are recovered. 

Both factorized expressions can be seen as approximations of the unknown, \emph{true} theory, up to corrections expressed as ratios of the relevant scales (power corrections, in the following).
In TMD factorization the power corrections scale as a power of $q_T/Q$, whereas in collinear factorization they scale as a power of $m/q_T$, up to further suppressed nonperturbative contributions~\cite{Collins:2011zzd}. 
We simply implement an estimate of these uncertainties in the well-known formula of an inverse-error weighting --or inverse-variance weighted average-- of two measurements to obtain our matched predictions. 
As such, it also automatically returns an evaluation of the corresponding matching uncertainty.

The method we propose can straightforwardly be extended to match any (un)polarized cross section differential in other variables, including for instance event shapes, multi-differential measurements or double parton scattering with a measured transverse momentum~\cite{Buffing:2017mqm}.

This paper is organized as follows:
in Sec.~\ref{s:average} we describe both factorization theorems for low and high transverse momenta, and how they are combined with the inverse-error-weighting method.  
In Sec.~\ref{s:implementation} we show through several examples ($Z$, $W$, $H^0$ and Drell-Yan lepton-pair production) how the method works.  
In Sec.~\ref{s:comparison} we compare the numerical results to the \icss subtraction scheme. 
Finally, Sec.~\ref{s:conclusions} gathers the conclusions and briefly discusses the applicability of our method to other processes.

\section{The Inverse-Error Weighting Method}
\label{s:average}

The main idea behind the scheme we are proposing is to use the power corrections to the involved factorization theorems in order to directly determine to which extent the approximations can be trusted in different kinematic regions, and to use this in order to bridge the intermediate region obtaining the complete spectrum. 
In this context, an inverse-error weighting is conceptually the simplest method one could think of.

Let us have a closer look at the TMD and collinear factorization theorems and their regions of validity, by considering a cross section $d\sigma$ differential in at least the transverse momentum $q_\st$ of an observed particle. 
For $q_\st \ll Q$, the TMD factorization can reliably be applied and the $q_\st$-differential cross section can generically be written as
\begin{align}
\label{e:approximationsTMD}
	d\s(q_\st,Q)\Big|_{q_\st \ll Q} &=
	{\cal W}(q_\st,Q) + \Bigg[{\cal O}\bigg(\frac{q_\st}{Q}\bigg)^a + {\cal O}\bigg(\frac{m}{Q}\bigg)^{a'} \Bigg]d\s(q_\st,Q) 
	\,, 
\end{align}
where $\cal{W}$ is the TMD approximation of the cross section $d\s$, the scale $m$ is a hadronic mass scale on the order of 1 GeV and $Q$ is the hard scale in the process, for instance the invariant mass of the produced particle. 
As $q_\st$ increases, the accuracy of the TMD approximation decreases and the power corrections are increasingly relevant until the expansion breaks down as $q_\st$ approaches $Q$.

On the contrary, for large $q_\st\sim Q \gg m$, the collinear factorization theorem applies and the $q_\st$-differential cross section can generically be written as
\begin{align}
\label{e:approximationsCOL}
d\s(q_\st,Q)\Big|_{q_\st\sim Q \gg m} &=
{\cal Z}(q_\st,Q) + {\cal O}\bigg(\frac{m}{q_\st}\bigg)^b 
d\s(q_\st,Q)
\,,
\end{align}
where $\cal Z$ is the collinear approximation of the full cross section $d\sigma$.
$\cal Z$ is calculated at a fixed-order in the strong coupling constant $\alpha_s$. 
For $q_\st\sim Q \gg m$, $\cal Z$ is a good approximation of the full cross section, but as $q_\st$ decreases the accuracy of the collinear approximation diminishes, which finally breaks down as $q_T$ approaches $m$. 

Armed with both these factorization theorems, valid in different and (sometimes) overlapping regions, the full $q_\st$ spectrum can be constructed through a matching scheme. 
Such a scheme must make sure that the result agrees with $\cal{W}$ in the small $q_\st$ region and with $\cal{Z}$ in the large $q_\st$ region, and that there is a smooth transition in the intermediate region.

As announced, in this paper we introduce a new scheme, the \emph{inverse-error weighting} (\inew\ for short), where the power corrections to the factorization theorems are used to quantify the trustworthiness associated to the respective contributions, and thus employed to build a weighted average. 
The resulting matched differential cross section over the full range in $q_\st$ is given by
\begin{equation}
\label{e:wav_sigma}
\overline{d\sigma}(q_\st,Q) 
= 
\omega_1 {\cal W}(q_\st,Q) + \omega_2 {\cal Z}(q_\st,Q) 
\,, 
\end{equation}
where the normalized weights for each of the two terms are
\begin{align}
	\omega_1 & = \frac{\Delta \mcal{W}^{-2}}{\Delta \mcal{W}^{-2} + \Delta \mcal{Z}^{-2} }\,, 
	& \omega_2 & = \frac{\Delta \mcal{Z}^{-2}}{\Delta \mcal{W}^{-2} + \Delta \mcal{Z}^{-2} }\,,
\end{align}
with $\D{\cal W}$ and $\D{\cal Z}$ being the uncertainties of both factorization theorems generated by their power corrections.
The uncertainty on the matched cross section simply follows from the propagation of 
these (uncorrelated) theory uncertainties:
\begin{align}
\label{e:err_wav_sigma}
\D\overline{d\s} &=
\frac{1}{ \sqrt{ \D \mcal{W}^{-2}+\D \mcal{Z}^{-2} } } 
= \frac{\D_{\cal W} \D_{\cal Z}}{\sqrt{\D_{\cal W}^2+\D_{\cal Z}^2}} \, d\s
\approx
\frac{\D_{\cal W} \D_{\cal Z}}{\sqrt{\D_{\cal W}^2+\D_{\cal Z}^2}} \, \overline{d\s}
\,,
\end{align}
where $\{\D{\cal W},\D{\cal Z}\} = \{\D_{\cal W},\D_{\cal Z}\}d\s$, and in the last step we have replaced the unknown true cross section $d\s$ by its estimated value $\overline{d\s}$.
We emphasize that the uncertainty on the matched cross-section, $\D\overline{d\s}$, is due \emph{only} to the matching procedure, which in the \inew\ method comes from the power-corrections.
In any phenomenological application one should also include, once the matched cross-section is obtained, all other sources of uncertainty, i.e. the ones related to higher perturbative orders and nonperturbative contributions.

Following Eqs.~\eqref{e:approximationsTMD} and~\eqref{e:approximationsCOL}, we numerically implement the uncertainties $\D_{\cal W}$ and $\D_{\cal Z}$ as
\begin{equation}
\label{e:num_errors}
\D_{\cal W} = \bigg(\frac{q_\st}{Q}\bigg)^{a} + \bigg(\frac{m}{Q}\bigg)^{a}
\,, 
\qquad
\D_{\cal Z} = \bigg(\frac{m}{q_\st}\bigg)^{b} \Bigg(1+ \ln^2\le(\frac{m_\st}{q_\st}\ri)\Bigg)
\,.
\end{equation}
As an Ansatz, we have taken $a=a^\prime$ and will discuss the impact of this choice at the end of Sec.~\ref{s:implementation}. 
In the region where $q_\st$ becomes different from $Q$, large logarithms will reduce the accuracy of the power counting which was done in the $q_\st \sim Q$ region.
We have thus included a $\ln^2(m_\st/q_\st)$ in $\D_{\cal Z}$, where the transverse mass is defined as $m_\st = \sqrt{Q^2 + q_\st^2}$, which is the expected typical leading logarithm in the fixed-order calculations.
This logarithm then allows us to have a more reliable estimation of the power corrections to the collinear result in the whole $q_\st$ range, and not only at $q_\st\sim Q$.

The values of the exponents $a$ and $b$ are given by the strength of the power corrections and depend on the details of the process and its factorization.
In the case of unpolarized processes, the smallest values allowed by Lorentz symmetry are $a=2$ and $b=2$, since $q_\st$ is the only transverse vector that explicitly appears in the factorization theorems.
This is consistent with what is found in Refs.~\cite{Balitsky:2017flc,Balitsky:2017gis} for the TMD factorization theorem, and in Refs.~\cite{Qiu:1990xxa,Qiu:1990xy} for the $q_\st$-integrated collinear factorization theorem, which should also apply for the $q_\st$-unintegrated when $q_\st\sim Q$.
We thus take $a=2$ and $b=2$ as the default choice for the numerical implementations.

In order to obtain a more conservative estimation of the power corrections in the presence of large logarithmic corrections, the values of $a$ and $b$ could be reduced (see Sec.~13.12 in Ref.~\cite{Collins:2011zzd}).
Moreover, smaller values are expected for spin-asymmetry observables, where $q_\st$ is not the only explicit vector, but also a transverse spin vector $S_\st$ contributes to the cross section.
Even though $a=1$ and $b=1$ might be an extreme choice, we have considered it to get first indications on the matching uncertainty in the polarized cases, which we plan to study in more detail in forthcoming publications.

Summarizing, we obtain the differential cross section for the full $q_\st$ spectrum as the \emph{weighted average}, Eq.~\eqref{e:wav_sigma}, of the TMD and collinear approximations $\cal W$ and $\cal Z$ with their weights calculated as the inverse of the square of the power corrections to the factorized expressions, as in Eq.~\eqref{e:num_errors}. 
The uncertainty of the matched result automatically follows from Eq.~\eqref{e:err_wav_sigma}.

Let us note that the derivation of the power corrections in both factorization theorems is only valid in and around their regions of validity. 
For example, for $q_\st > Q$ the power counting leading to the power corrections for the TMD cross section breaks down. 
In this region, however, the collinear-factorization theorem fully dominates the result and the matched result correctly reproduces the $\cal Z$-term and thereby the cross section (an analogous logic applies to small $q_\st$).

\section{Illustration of the method}
\label{s:implementation}

\begin{figure*}[hbt!]
\begin{center}
\includegraphics[width=0.33\textwidth]{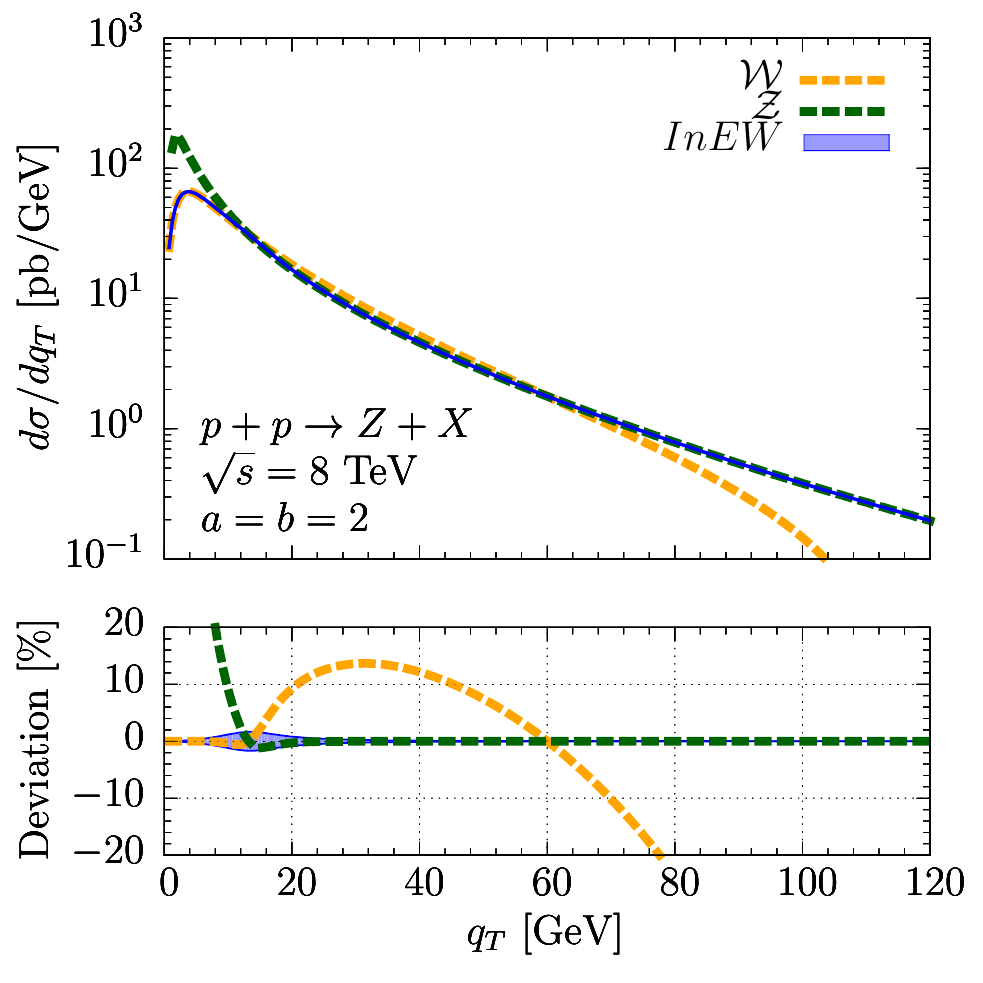}
\includegraphics[width=0.33\textwidth]{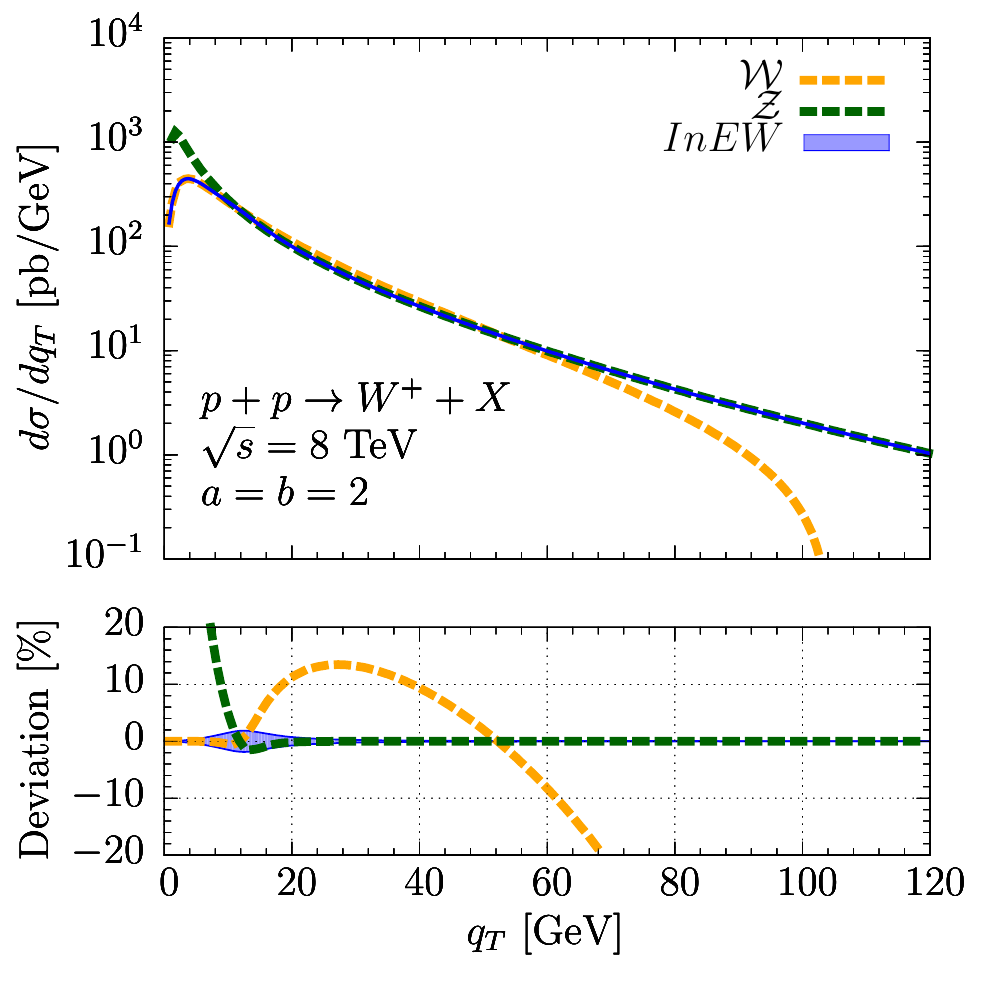}
\includegraphics[width=0.33\textwidth]{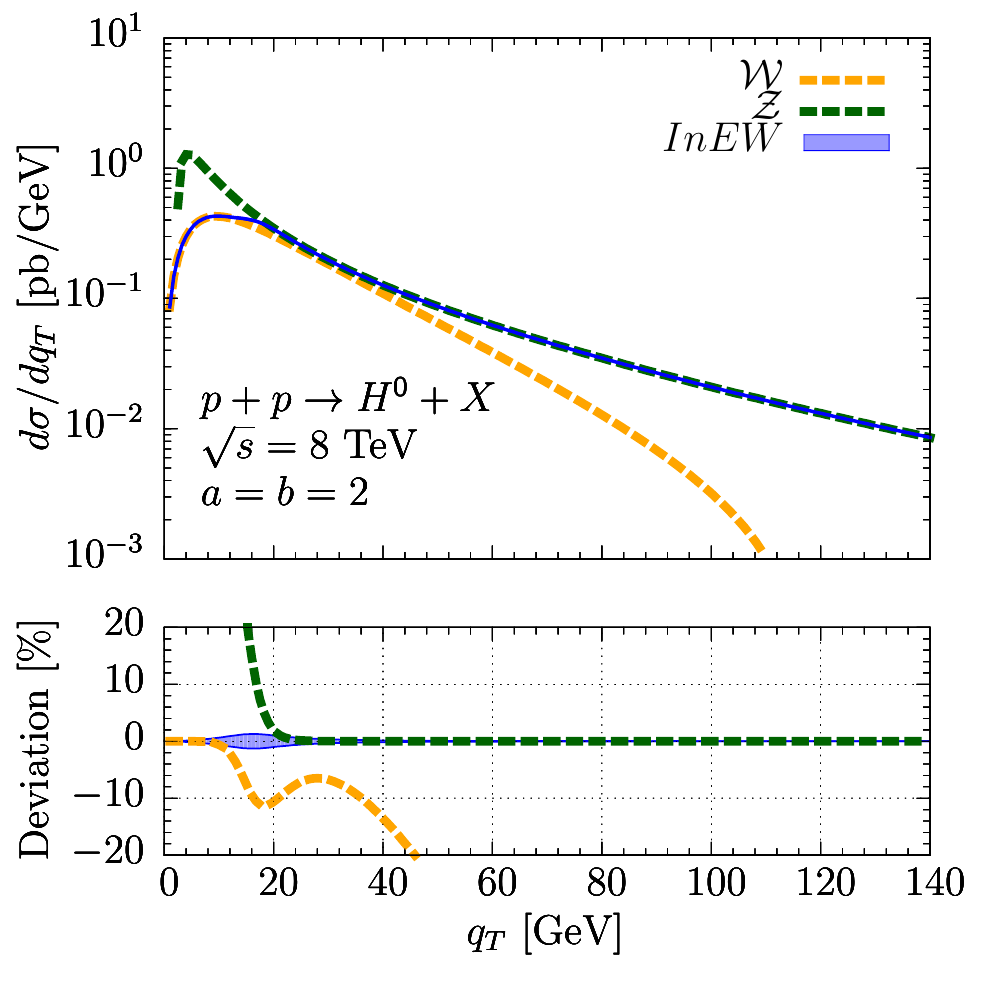}
\\
\includegraphics[width=0.33\textwidth]{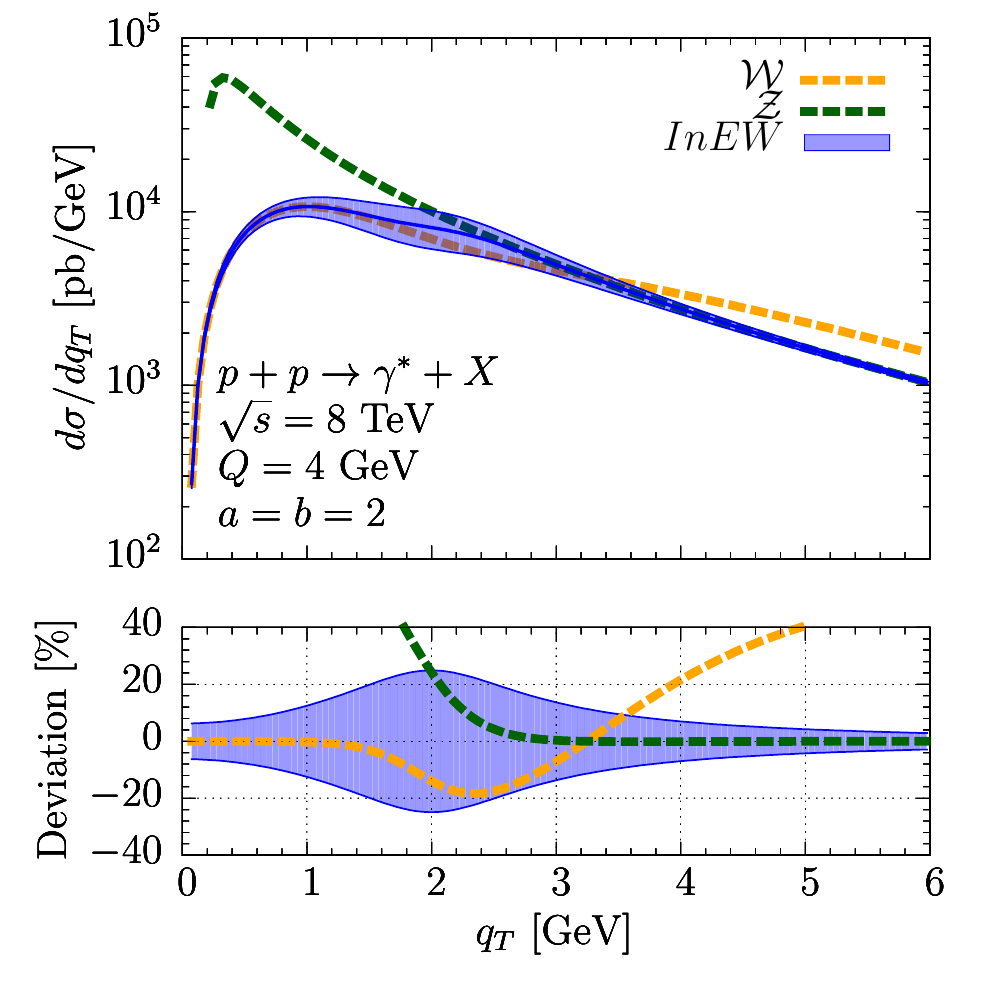}
\includegraphics[width=0.33\textwidth]{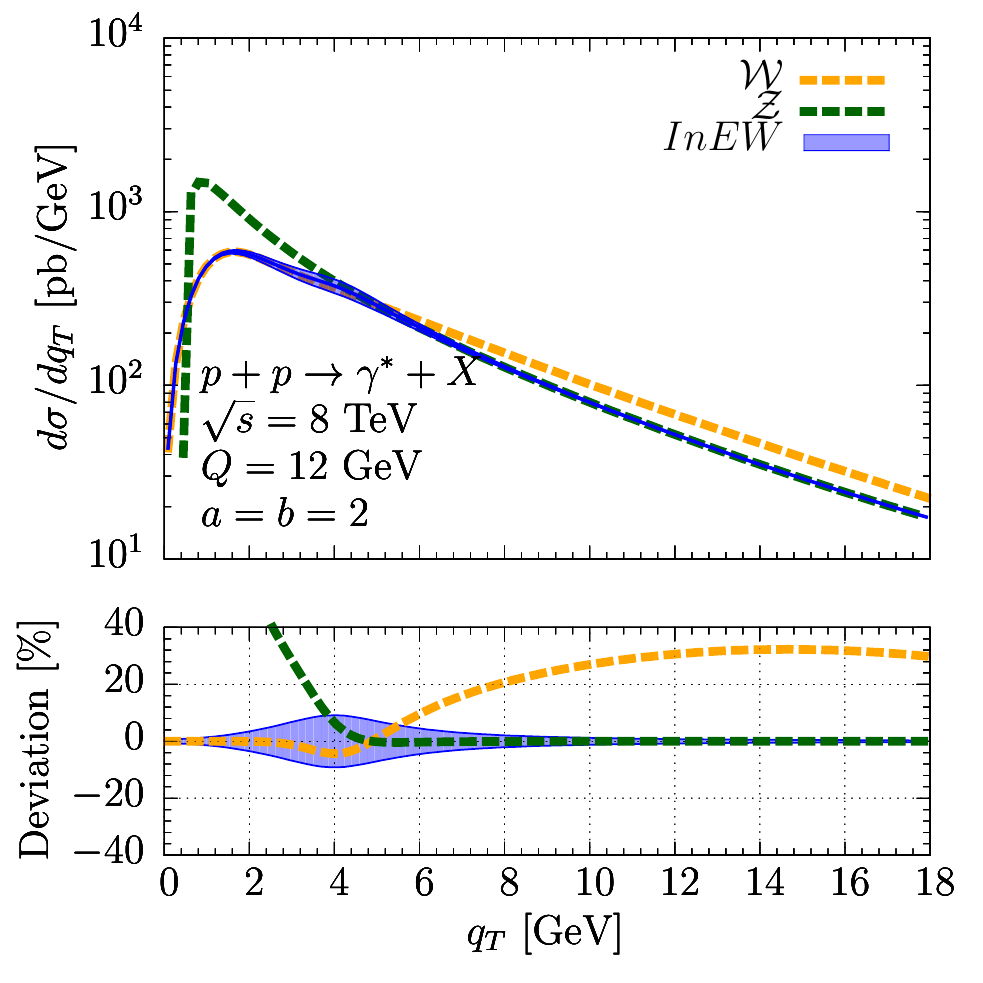}
\includegraphics[width=0.33\textwidth]{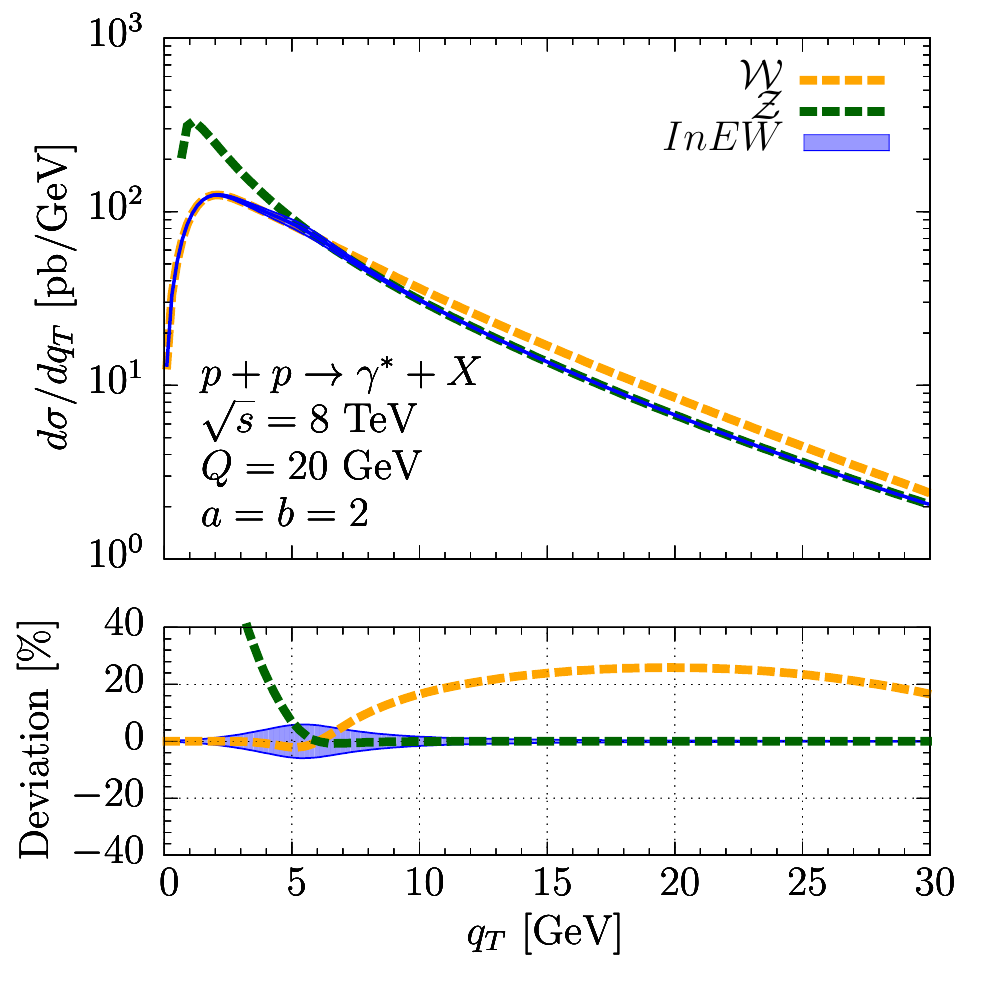}
\end{center}
\caption{
The resummed term $\cal W$ (yellow curve), the fixed-order term $\cal Z$ (green curve), and the matched cross section in the \inew\ approach (blue band) for 
$Z$ boson production (top-left), 
$W^+$ boson production (top-center), 
$H^0$ boson production (top-right), 
Drell-Yan lepton-pair production with $Q=4$~GeV (bottom-left),  
$Q=12$~GeV (bottom-center), and 
$Q=20$~GeV (bottom-right).   
All processes are initiated by proton-proton collisions with $\sqrt{s}=8$~TeV. 
The uncertainty on the matched cross section is only due to the matching scheme, i.e. including power-correction uncertainties, and no other effects are added, such as the perturbative-scale variations and the nonperturbative contributions.
Lower panels quantify the deviation of the $\cal W$- and $\cal Z$-terms with respect to the matched cross section, as well as its matching uncertainty.}
\label{f:xsections}
\end{figure*}

In the following, we illustrate how the method works for the computation of the $q_\st$ distribution of different electroweak bosons produced in proton-proton collisions at the LHC at $\sqrt{s}=8$~TeV. 
In particular, we will consider the following processes:
\begin{itemize}
\item $Z/W$ boson production (Sec.~\ref{ss:ZW_prod})
\item Drell-Yan (DY) lepton-pair production (Sec.~\ref{ss:DY})
\item $H^0$ boson production (Sec.~\ref{ss:H_prod}).
\end{itemize}  
These processes are sensitive to either quark TMDs ($Z/W$ boson and DY production) or gluon TMDs ($H^0$ boson production), and allow us to illustrate the implementation of the matching scheme from low to high values of the hard scale. 

The cross sections differential with respect to the transverse momentum $q_\st$ of $Z/W$ boson and Drell-Yan production have been computed using the public code \dyqt\footnote{\href{http://pcteserver.mi.infn.it/~ferrera/research.html}{http://pcteserver.mi.infn.it/$\sim$ferrera/research.html}}~\cite{Bozzi:2008bb,Bozzi:2010xn}. 
For $H^0$ boson production we have used the public code \hqt\footnote{\href{http://theory.fi.infn.it/grazzini/codes.html}{http://theory.fi.infn.it/grazzini/codes.html}}~\cite{deFlorian:2011xf}.

We have worked with the highest perturbative accuracy implemented in \dyqt\ and \hqt :   
NNLL (next-to-next-to-leading logarithmic) accuracy in the resummed contribution $\cal W$ (i.e. $\Gamma_{\text{cusp}} \sim {\cal O}(\alpha_s^3)$) and
NLO (next-to-leading order) corrections (i.e. ${\cal O}(\alpha_s^2)$) at large $q_\st$ for the fixed-order contribution $\cal Z$.
For collinear PDFs we have used the NNPDF3.0 set at NNLO with $\as(M_Z)=0.118$~\cite{Ball:2014uwa}.

The treatment of the different $b_\st$ regions (where $b_\st$ is the Fourier-conjugated variable to the observed transverse momentum $q_\st$) is identical in both \hqt\ and \dyqt. 
The large $b_\st$ region is treated with the so-called {\em complex $b_\st$} (or {\em minimal}) prescription, which avoids the Landau pole in the coupling constant by deforming the integration contour in the complex plane~\cite{Laenen:2000de,Kulesza:2002rh}.
The small $b_\st$ region, instead, is treated by replacing the $\log (Q^2b_\st^2)$ with $\log(Q^2b_\st^2 + 1)$~\cite{Parisi:1979se,Bozzi:2005wk}, avoiding unjustified higher-order contributions.
This is analogous to introducing a lower cutoff $b_{\text{min}}$ in $b_\st$ space~\cite{Collins:2016hqq,Signori:2016lvd,Boer:2014tka,Boer:2015uqa,Collins:2016hqq}. 
We note that this cutoff is crucial in order to recover the integrated collinear factorization result upon integration over the transverse momentum.

In \dyqt, the nonperturbative TMD part in the resummed term 
is implemented as a simple Gaussian smearing factor in $b_\st$ space~\cite{Bozzi:2008bb,Bozzi:2010xn}
$
\exp(S_{NP}) = \exp( -g_{NP}\, b_\st^2)
$.
%
Since we are interested in processes at different energy scales, we have included a logarithmic dependence of $g_{NP}$ on the invariant mass $Q$ of the produced state (see e.g. Ref.~\cite{Bacchetta:2015ora}) to mimic more realistic values: 
$
g_{NP}(Q) =
g_{NP}^0\, \ln({Q^2}/{Q_0^2})
$
with $Q_0=1$ GeV.
Thus, we can write
$
g_{NP}(Q) =
g_{NP}(M_Z) {\ln\big(Q^2 \big)}/{\ln\big(M_Z^2 \big)}$.
In \hqt\ an analogous smearing factor was introduced.
For the gluon TMDs there is significantly less experimental input and thus phenomenological information (see however Ref.~\cite{Lansberg:2017dzg}) and we have rescaled the nonperturbative parameter for quark TMDs by a Casimir scaling factor $C_A/C_F$ (see Sec.~\ref{ss:H_prod}), where $C_A = N_c$, $C_F = (N_c^2-1)/2N_c$ and $N_c=3$ is the number of colors.
Let us however note that such nonperturbative factors, which would be essential for
a proper comparison with data, are not involved in the matching procedure.

\dyqt\ and \hqt\ allowed us to separately compute the cross section at low $q_\st$ ($\cal W$) and at high $q_\st$ ($\cal Z$) from which we have implemented the matching following our \inew\ method. 
These codes also allowed us to compute the asymptotic limit~\cite{Arnold:1990yk,Collins:1984kg,Collins:2016hqq} of the resummed contribution ($\cal A$) which we will use for the comparison with \icss\ method.
 
The uncertainties in the following sections will purely be from the \inew\ matching scheme, namely induced by the estimation of the power corrections. 
Additional uncertainties due to scale variations, collinear-parton distributions and TMD nonperturbative uncertainties should be added for a fair comparison with data. 
We stress that this remark would apply to any (un)matched computations.
We leave for a future publication the phenomenological study of the \inew\ scheme, where the uncertainties on the functional form and the parameters of the nonperturbative contribution will be considered (see e.g. Refs.~\cite{Scimemi:2017etj,Bacchetta:2017gcc} for recent phenomenological works).

\subsection{$Z/W$ boson production}
\label{ss:ZW_prod}

In this section we study $Z/W$ boson production.
We work in the narrow width approximation and include the branching ratio into two leptons~\cite{Bozzi:2008bb,Bozzi:2010xn}.

In Fig.~\ref{f:xsections} (top-left) we show the full transverse-momentum spectrum calculated with our \inew\ matching of the $\cal W$ and $\cal Z$-terms for $Z$-boson production.
The nonperturbative parameter we used is $g_{NP}(M_Z)=0.8~\gev^{2}$ \cite{Kulesza:2002rh}.
The central curve corresponds to $\overline{d\sigma}$ and the band to its variation by $\pm \Delta \overline{d\sigma}$ (see Eq.~\eqref{e:wav_sigma} and Eq.~\eqref{e:err_wav_sigma}). We also show the $\cal W$- and $\cal Z$-terms individually. 
The lower panels in Fig.~\ref{f:xsections} quantify the deviation\footnote{
By \emph{deviation} we mean the percentage difference between two given curves.
For the $\cal W$-term we plot $100\!\cdot\!({\cal W} - \overline{d\s})/\overline{d\s}$, and similarly for the rest.} 
of the $\cal W$- and $\cal Z$-terms with respect to the matched cross section, as well as its matching uncertainty. 

The $\cal Z$-term is ill behaved towards small values of the transverse momentum due to the presence of the large logarithms in $Q/q_\st$, while the $\cal W$-term tends towards negative values for large $q_\st$. 
There is a quite broad intermediate region where both results are similar, and where both factorization theorems are on relatively stable ground. 
This makes the matching between the two theorems particularly simple, and well behaved.

The cross section matched in the \inew\ scheme follows the resummed $\cal W$-term up to $q_\st \sim 15$ GeV and then approaches the fixed-order $\cal Z$-term. 
The uncertainty from the power corrections is small in the large and very small $q_\st$ regions, but increases in the region around the value of $q_\st$ where $\D_{\cal W}=\D_{\cal Z}$ (i.e. where both weights are close to $\tfrac{1}{2}$).

The results for $W^+$ production are shown in Fig.~\ref{f:xsections} (top-center). 
The scale-dependent nonperturbative parameter is modified to $g_{NP}(M_W)\simeq 0.78~\gev^{2}$ by the change of the hard scale to the mass of the $W$ boson ($M_W=80.385$ GeV). 
The results for the matched cross section closely resemble those for the $Z$ boson, which is to be expected since both processes have a similar hard scale and probe quark and antiquark distributions. 
The transition point between the $\cal W$-term and the $\cal Z$-term has moved down to slightly lower $q_\st$, and the uncertainty is a little larger. The result for $W^-$ production is very similar, with just a different normalization for the differential cross section.

\subsection{Drell-Yan process}
\label{ss:DY}

In this section we study Drell-Yan lepton-pair production, or more precisely, virtual-photon ($\g^\star$) production.
The nonperturbative parameters are now given by
$g_{NP}(4 \text{ GeV})\simeq 0.25~\gev^{2}$, 
$g_{NP}(12 \text{ GeV})\simeq 0.44~\gev^{2}$ and
$g_{NP}(20 \text{ GeV})\simeq 0.53~\gev^{2}$. 
The results for the matched cross section for DY production are shown for the invariant masses $Q=4,12,20$~GeV in Fig.~\ref{f:xsections} (bottom). 
The values are chosen to complement the results for the heavy vector-boson and $H^0$ boson cross sections, and to demonstrate how the method performs at different scales.

Let us start our discussion from the lowest scale, $Q = 4$~GeV. 
This value is chosen to demonstrate what happens when the hard scale is very low, and when the intermediate region, where both TMD and collinear factorizations are valid, collapses. 
The matched cross section follows the TMD result up to larger fractions of $Q$ than it did for heavy vector-boson production, starting to tend towards the collinear result around $q_\st\sim Q/2$.
For such low scales, power corrections are of course likely to be large. 
This is nicely reflected by the uncertainty band of the \inew\ matched result which reaches maximum values of around 30\%. 
We note that significantly lowering the center of mass energy does not change the qualitative discussion of the matching method. 

Increasing the invariant mass of the produced boson, the uncertainty of the \inew\ scheme decreases and the transition between the two factorization theorems moves towards smaller fractions of $q_\st / Q$. 
The region where the results of both theorems are relevant also occupies a smaller and smaller portion of the $q_\st$ spectrum. 
At $Q=12$~GeV, the maximal uncertainty has decreased below 20\% and, at $Q=20$~GeV, is less than 10\%.

\subsection{$H^0$ boson production}
\label{ss:H_prod}

In this section we study $H^0$ boson production.
The heavy-top effective theory is used to integrate out the top quark, resulting in a direct coupling between gluons and the $H^0$ boson.

Unlike the previous processes, $H^0$ production directly probes gluon TMDs (see e.g. Refs.~\cite{Boer:2010zf,Qiu:2011ai,Boer:2011kf,Ma:2012hh,Boer:2012bt,Boer:2014lka,Dunnen:2014eta,Echevarria:2015uaa,Dumitru:2015gaa,Boer:2016xqr,Boer:2016bfj,Boer:2016fqd,Lansberg:2017tlc,Lansberg:2017dzg}). 
There is much less phenomenology and therefore knowledge about gluon TMDs than for quarks. 
As already mentioned in Sec.~\ref{s:implementation}, in order to obtain a reasonable value for the nonperturbative parameter we use Casimir scaling. 
This results in
$g_{NP}^g(125 \text{ GeV}) =$\\ 
$(C_A/C_F)g_{NP}(125 \text{ GeV}) \simeq 1.93~\gev^{2}$.

Fig.~\ref{f:xsections} (top-right) shows the matched cross section in the \inew\ scheme. 
It follows the $\cal W$-term up to $q_\st\sim15$~GeV and then approaches the $\cal Z$-term. 
The uncertainty band is narrow, as power corrections are strongly suppressed in the entire spectrum. 

The small size of the power corrections in combination with the large difference between the two factorized approximations of the cross section is a challenge for the matching in the intermediate region. 
At $q_\st\sim 15$~GeV, the power corrections $\Delta_{\cal W}$ and $\Delta_{\cal Z}$ are both below 0.05, but the $\cal Z$-term is 50\% larger than the $\cal W$. 
This is, however, no longer surprising when taking into account the large uncertainty associated to the $H^0$ boson transverse-momentum spectrum coming from the scale variations~\cite{Neill:2015roa}. 
It is therefore likely that higher-order corrections will bring the collinear and TMD results closer to each other, resulting in a smoother matching.

Finally, let us note that we did not observe any relevant variations of the central value of the matched cross section when lowering the exponents from $a=b=2$ to $a=b=1$.
However, as expected, the matching uncertainty significantly grows. 
For $Z$, $W^+$ and $H^0$ boson production cases, the uncertainty at its maximum is inflated 7-8 times, reaching $\sim15\%$ at $q_\st\sim 15$~GeV and remaining larger than 5\% from roughly 4 to 40 GeV. 
For the Drell-Yan case, whose transverse-spin-asymmetry study is a hot topic within the TMD community, the uncertainty rather inflates by a factor of 2 to 3 depending on the lepton-pair mass. 

On the other hand, the matching is quite stable under variations of the exponent $a'$ compared to $a$. 
For $a=a'$, the uncertainty associated with $a$ dominates down to very low $q_\st\sim$~1 GeV, and therefore dominates the region where both TMD and collinear results are relevant.
Lowering $a'$ leads to a slightly larger uncertainty in the low $q_\st$ region and can, for low $Q$ Drell-Yan, shift the transition between the TMD and collinear results towards slightly lower $q_\st$. 
However, we would like to note here that the exponents $a$, $a'$ and $b$ can be fixed for a given process by the order at which the different power corrections contribute.

\section{Comparison to CSS subtraction}
\label{s:comparison}

\begin{figure*}[hbt!]
\begin{center}
\includegraphics[width=0.33\textwidth]{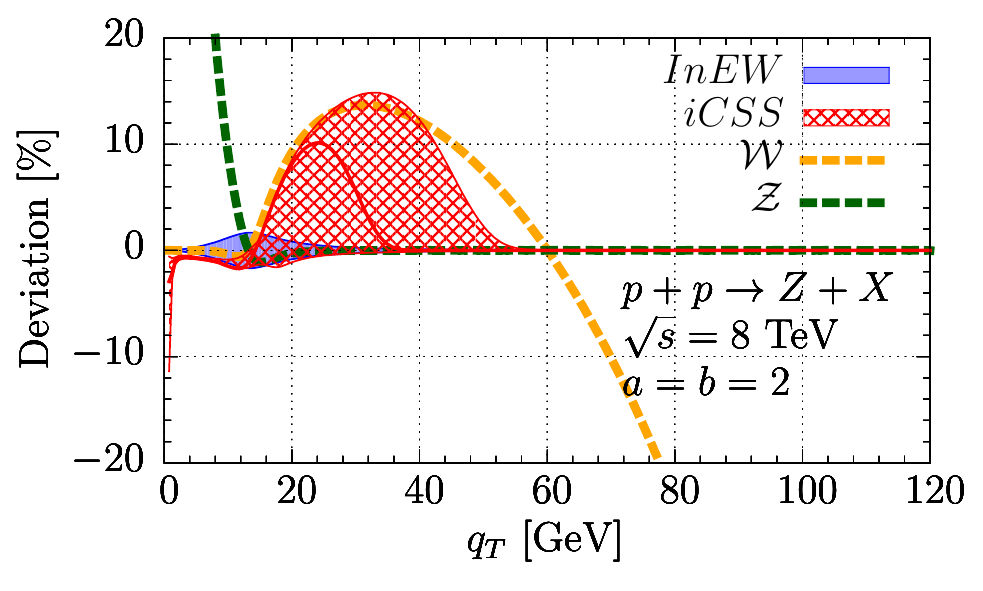}
\includegraphics[width=0.33\textwidth]{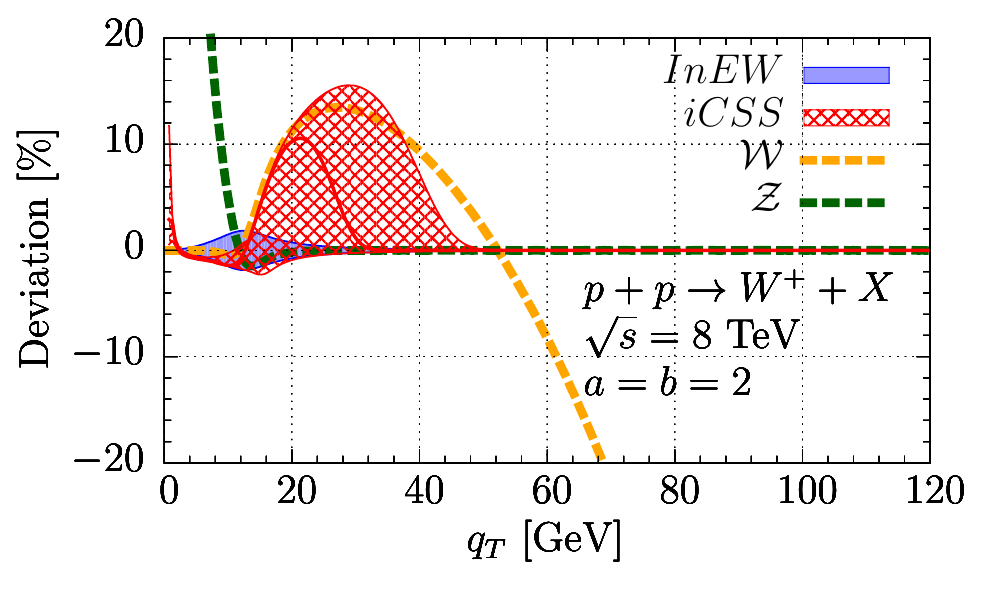}
\includegraphics[width=0.33\textwidth]{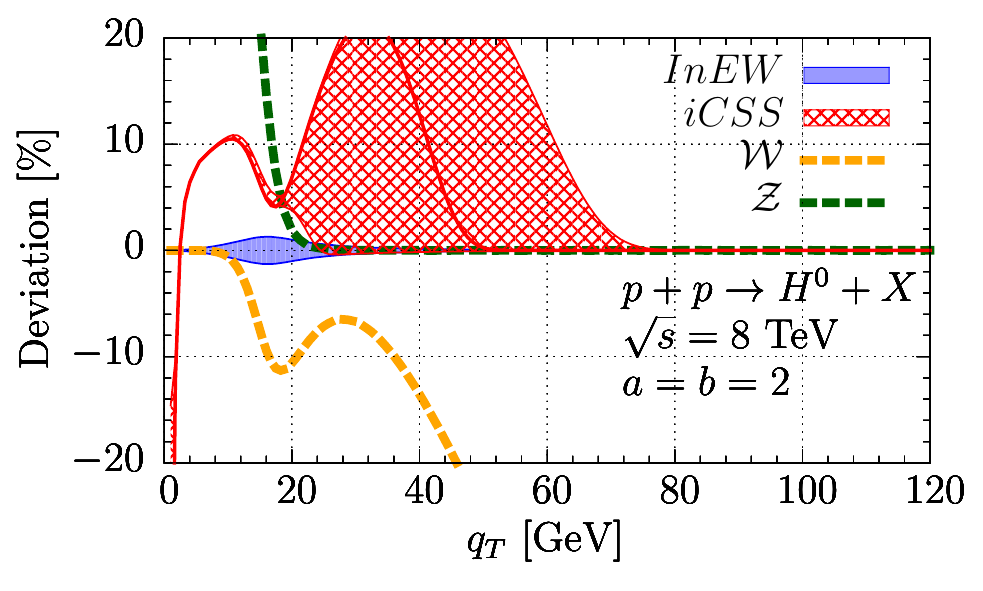}
\\
\includegraphics[width=0.33\textwidth]{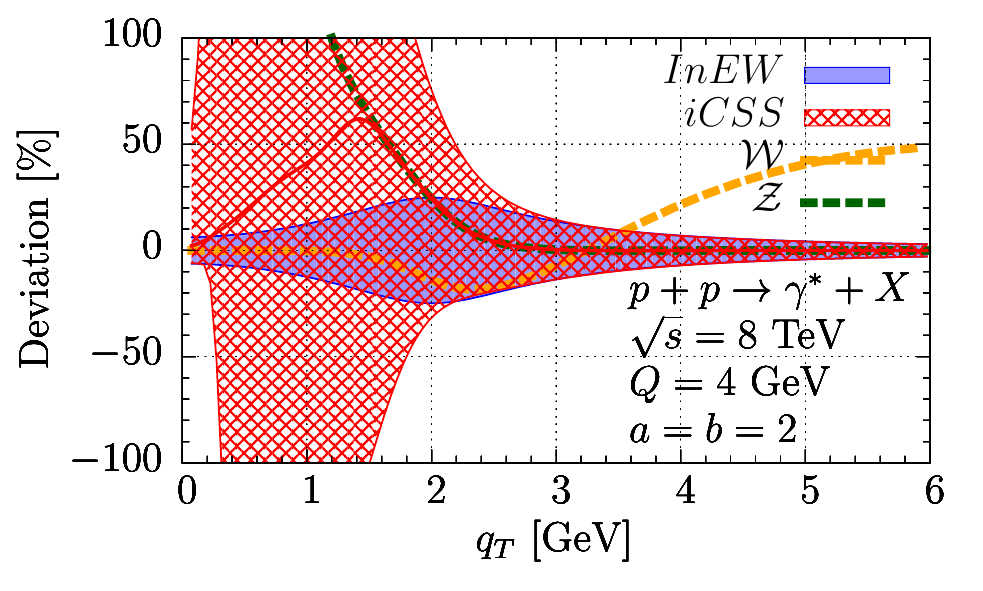}
\includegraphics[width=0.33\textwidth]{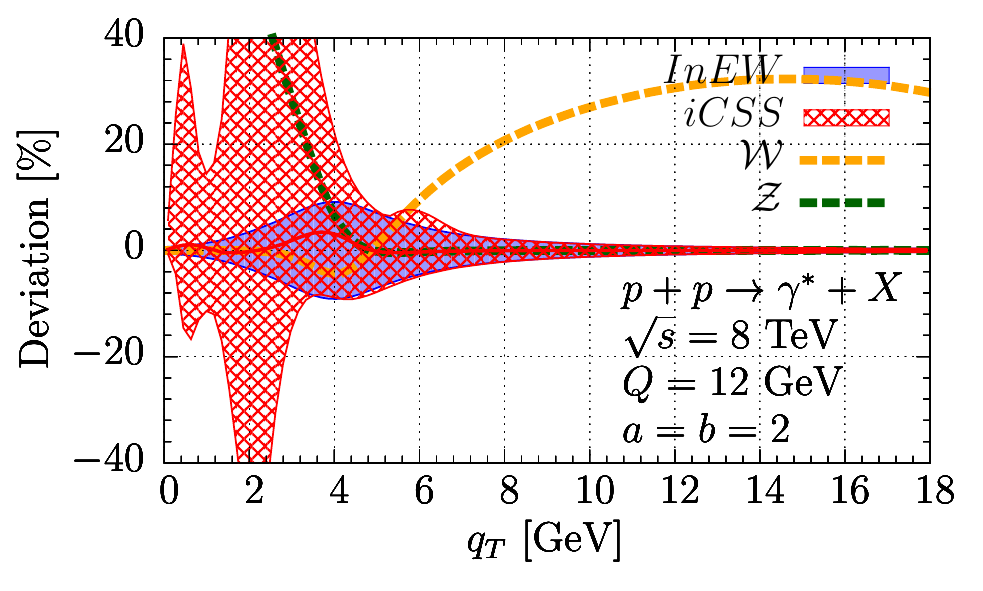}
\includegraphics[width=0.33\textwidth]{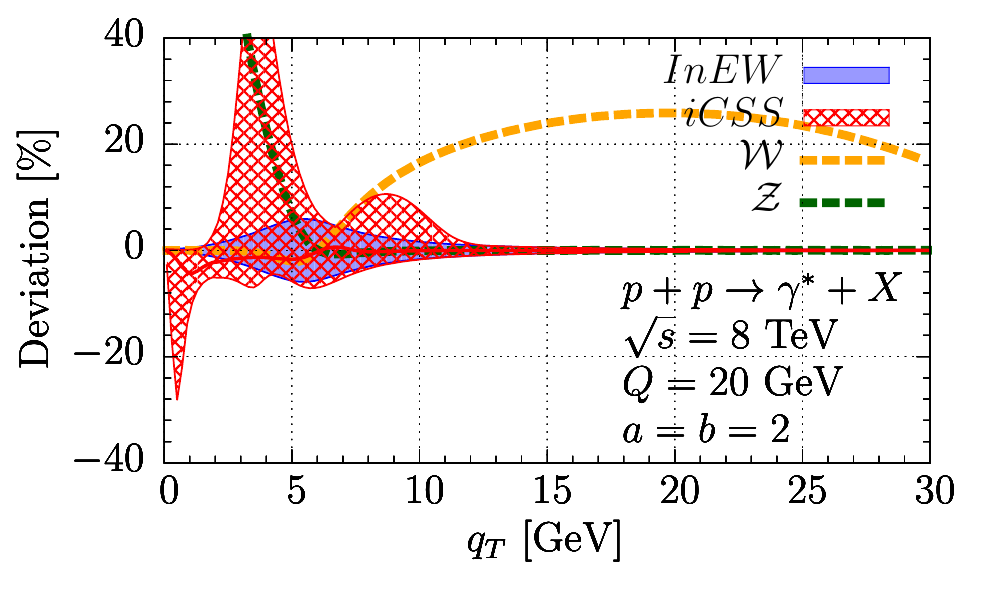}
\end{center}
\caption{
From left to right and top to bottom, comparison between the \inew\ and the \icss\ schemes for: $Z$ boson production, $W^+$ boson production, $H^0$ boson production and Drell-Yan lepton-pair production at $Q=4,12,20$ GeV.
}
\label{f:compare}
\end{figure*}

In this section, we compare the matched-cross-section results in the \inew\ scheme with the results in the \icss subtraction scheme of Ref.~\cite{Collins:2016hqq}. 
We therefore briefly introduce the features of the \icss\ method which are of relevance for our comparison, and refer to Ref.~\cite{Collins:2016hqq} for a more detailed discussion.

The widely used CSS method~\cite{Collins:1981uk,Collins:1981uw,Collins:1984kg,Arnold:1990yk} allows for a matching of the TMD result ($\cal W$) and the fixed-order result ($\cal Z$) in an additive way. 
Double counting is avoided by the subtraction of the asymptotic term ($\cal A$), i.e. the fixed-order expansion of the perturbative result of $\cal W$. 
For applications of the method in processes with a low hard-scale, see, e.g., Ref.~\cite{Boglione:2014oea} for Semi-Inclusive Deep-Inelastic Scattering (SIDIS) and Chap. 8 in Ref.~\cite{Signori:2016lvd} for $\eta_{b}$ production in proton-proton collisions. 
Applications in processes with a higher hard scale can be found in, e.g., Refs.~\cite{Catani:2015vma,Becher:2011xn,Becher:2012yn,Signori:2016lvd}. 

The method, although successful, runs into difficulties at small $q_\st$, due to incomplete cancellations between the fixed-order and the asymptotic results, and also at large $q_\st$, due to incomplete cancellations between the resummed and the asymptotic results.
At low $q_\st$ the problems are especially manifest when the hard scale Q is not large, namely when there is little or no overlap between the regions where the TMD and collinear factorization theorems are valid~\cite{Boglione:2014oea,Collins:2016hqq}.

Recently, a solution to these issues has been proposed in Ref.~\cite{Collins:2016hqq}, the \icss\ method. 
In order to enforce the required cancellations, the different terms in the cross section are multiplied by cutoff functions, damping them outside their region of validity.
This solves the problem of the incomplete cancellations, but introduces a dependence both on the functional form of the cutoff functions and on the point in $q_\st$ where one switches on and off the different contributions. 

The cross section in the \icss\ method is written as 
\begin{align}
\label{e:app_iWY}
d\sigma(q_\st,Q) &= {\cal W}_{\text{\icss}}(q_\st,Q) + {\cal Y}_{\text{\icss}}(q_\st,Q) \ ,
\end{align}
where 
\begin{align}
\label{e:WZAnew}
{\cal W}_{\text{\icss}}(q_\st,Q) &= {\cal W}(q_\st,Q)\, \L^{\cal W}(q_\st,Q; \eta, r)
\,,
\nn\\
{\cal Y}_{\text{\icss}}(q_\st,Q) &= {\cal Z}_{\text{\icss}}(q_\st,Q) - {\cal A}_{\text{\icss}}(q_\st,Q)
\,,
\nn\\
{\cal Z}_{\text{\icss}}(q_\st,Q) &= {\cal Z}(q_\st,Q)\, \L^{\cal Z}(q_\st; \l, s)
\,,
\nn\\
{\cal A}_{\text{\icss}}(q_\st,Q) &= {\cal A}(q_\st,Q)\, \L^{\cal W}(q_\st,Q; \eta, r)\, \L^{\cal Z}(q_\st; \l, s)
\,,
\end{align}
with the cutoff functions
\begin{align}
\L^{\cal W}(q_\st,Q; \eta,r) = \exp \bigg\{ -\bigg( \frac{q_\st}{\eta Q} \bigg)^r \bigg\}
\,,
\nn\\
\L^{\cal Z}(q_\st;\lambda, s) = 
1 - \exp \bigg\{ -\bigg( \frac{q_\st}{\lambda} \bigg)^s \bigg\}
\,.
\end{align}
The parameters $\{\eta,\l\}$ control the value of $q_\st$ around which the cutoffs start, whereas the exponents $\{r,s\}$ control the steepness of these cutoffs\footnote{The authors of Ref.~\cite{Collins:2016hqq} also introduce a small-$b$ cutoff ($b_{\text{min}}$ prescription) in the $\cal W$-term, which has an effect as well in the way the asymptotic $\cal A$-term is calculated.}.
In simple terms, the damping function $\L^{\cal W}$ switches off both the $\cal W$-term and the $\cal A$-term at large $q_\st$, while the damping function $\L^{\cal Z}$ switches off both the $\cal Z$-term and the $\cal A$-term at small $q_\st$.
For intermediate $q_\st$, the three terms are kept.

The values for these four parameters given in Ref.~\cite{Collins:2016hqq} are 
$\{\eta,r\} = \{1/3, 8\}$ and $\{\l,s\} = \{2/3~\gev,4\}$.
We have chosen a different default value for $\lambda$ ($\lambda=1$~GeV) for switching off the $\cal Z$ and $\cal A$ towards low $q_\st$ values, in order for the cross section not to start deviating from the $\cal W$ towards too low $q_\st$. 
The variations of the parameters we perform however include also the default value of Ref.~\cite{Collins:2016hqq}. 

To be able to compare with the \icss\ approach we need to construct a way to estimate the matching uncertainty in the \icss\ scheme, both due to the power corrections and to the parameters in the matching scheme. 
To do so, we note that the cross section in the \icss\ method can be written as:
\begin{align}
\label{eq:piecewise}
d\s(q_\st,Q) &=
  \begin{cases}
  {\cal W} + \D_{\cal W}d\s &\,,\quad q_\st \lesssim \l\\
  {\cal W} + {\cal Z} - {\cal A} + \D_{\cal W}\D_{\cal Z}d\s &\,,\quad \l\lesssim q_\st \lesssim \eta Q\\
  {\cal Z} + \D_{\cal Z}d\s &\,,\quad q_\st \gtrsim \eta Q\\
  \end{cases}
  \ , 
\end{align}
since the damping functions $\L^{\cal W}$ and $\L^{\cal Z}$ are devised as (almost) step functions. 
At small $q_\st$, since the cross section is effectively given by the $\cal W$-term, the power counting (relative) error will be $\D_{\cal W}$ (see \eq{e:approximationsTMD}).
At large $q_\st$, the cross section is effectively given by the $\cal Z$-term, and the power counting (relative) error will be $\D_{\cal Z}$ (see \eq{e:approximationsCOL}).
In the intermediate region the cross section is given by the subtraction of the double-counted contributions, and thus the power counting (relative) error is $\D_{\cal W} \D_{\cal Z}$~\cite{Collins:2016hqq}.
We therefore estimate the error from subleading powers in the \icss\ method (as a function of $q_\st$) as
\begin{align}
\label{eq:iWYerror}
\frac{1}{d\s}\D d\s\Big|_{\icss} &=
\D_W \big[1-\L^{\cal Z}\big] + 
\D_W\D_Z \L^{\cal W}\L^{\cal Z} + 
\D_Z \big[1-\L^{\cal W}\big]
\,.
\end{align}

In addition to this uncertainty from the power corrections, we need to consider the uncertainty that comes from the variation of the matching parameters in the \icss\ approach\footnote{These should not be confused with the uncertainties from the perturbative-scale variations and the nonperturbative contributions.}.
In particular, we take the default values $\eta=1/3$ and $\l=1~\gev$ (different from the one proposed in Ref.~\cite{Collins:2016hqq}) and vary them by 50\%, i.e. $\eta\in[1/6,1/2]$ and $\l\in[0.5,1.5]~\gev$.  
We keep the exponents $\{r,s\}$ constant, since they have to be large enough to give almost step functions, and then their variation does not have any relevant impact.

In the intermediate region, this method has a potential advantage over the \inew\ in terms of the formal power counting uncertainty, i.e. $\D_{\cal W}\D_{\cal Z}/(\D_{\cal W}^2 + \D_{\cal Z}^2)^{1/2}$ for \inew\ compared to $\D_{\cal W}\D_{\cal Z}$ for \icss\ (where no variation of the matching parameters is included~\cite{Collins:2016hqq} though). 
This is of value, in particular, in high-scale processes such as $Z$ boson production, where there is an overlap region where the approximations in both of the two factorization theorems are appropriate. 
When the hard scale of the process is reduced, the overlap of the two factorization theorems decreases. 
As this happens, the subtraction method no longer benefits from the power counting advantage, since the uncertainty from the matching parameters is large, as we now demonstrate.

In Fig.~\ref{f:compare}, we show the numerical differences between the \inew\ and the \icss\ schemes for $Z$ boson production, $W^+$ boson production, $H^0$ boson production, and Drell-Yan lepton-pair production at $Q=4,12,20$ GeV.
The total uncertainty for the \icss\ approach shown in Fig.~\ref{f:compare} is obtained as the envelope of the uncertainty bands $d\s\pm\D d\s$, where each band corresponds to one of the mentioned choices of the matching parameters $\{\eta,\lambda\}$.
We again note that the uncertainties shown in Fig.~\ref{f:compare} are only due to the different matching schemes, and do not include other effects such as the perturbative-scale variations and the nonperturbative contributions, which are common to both.

Starting with the $Z$ and $W$ boson production and comparing the \inew\ results to those in the \icss\ scheme, we can notice that where the uncertainty in the \inew\ method is the largest, the \icss\ scheme produces a significantly smaller uncertainty. 
This is precisely due to the reduction of the power corrections obtained in the intermediate region when subtracting the asymptotic term $\cal A$. 
At the scale of the $Z$ boson mass, there is a significant overlap of the two regions where the two factorization theorems apply. 
However, we can also see that as we approach the regions of the matching points between the low and intermediate transverse momentum, or between the intermediate and high transverse momentum, the choice of the matching parameters has a large impact on the results. 
Unlike the \inew\ scheme, the \icss\ follows more closely the $\cal W$-term up to larger values of $q_\st$, but the extent to which this holds true has a strong dependence on the value of the largest matching point. 
This is clearly reflected in the size of the uncertainty in this region of transverse momentum.
For both processes, the uncertainty band for the \inew\ method is symmetric around the central value, while the estimation of the uncertainty for the \icss\ is asymmetric, originating mainly from the variation of the matching parameters. 

For DY at $Q=4$ GeV, the \icss\ scheme runs into difficulties. 
There is no space left for the intermediate region, and the matching points $\lambda$ and $\eta Q$ are very close to each other. 
This leads to a very large uncertainty. This is not surprising considering that the main advantage of the method is in the power counting uncertainty in the intermediate region. 
Moreover, for our choice of the default values for the parameters, the central curve in the \icss\ lies far away from the central curve in the \inew\ scheme at low and intermediate $q_\st$ values. 
The central curve in the \icss\ scheme moves from the resummed to the fixed-order result at a lower transverse momentum than the central curve in the \inew\ scheme, the opposite to what we could see in $Z/W$ boson production. 

Let us now compare the \inew\ and \icss\ schemes at $Q=12,20$~GeV, where there is more space for the intermediate region and the uncertainty in the \icss\ scheme improves. 
The \icss\ uncertainty at the larger transverse-momentum values is dominated by the variation of the matching point and remains of similar size regardless of the scale. 
A smaller (larger) variation of the associated parameter would of course lead to a smaller (larger) estimate of the associated uncertainty. 

For $H^0$ boson production, the advantage in the intermediate region of the \icss\ scheme is clearly visible, with a very small uncertainty band for low $q_\st$. 
The larger dependence on the choice of the upper matching point is however still present.  
Both schemes produce results which are clearly outside their uncertainty bands for a large range of intermediate transverse momenta. 
At this point, we emphasize that for $H^0$ production there is a large uncertainty coming from the scale variations~\cite{Neill:2015roa}. 
Therefore, the difference between the two methods will be drowned in the other uncertainties, given the currently available perturbative accuracy.
At very low $q_\st$ the \icss\ rapidly starts to deviate from the resummed calculation, but this is difficult to interpret. 
Changing the values of the  matching parameter associated with the transition between the low and intermediate region would fix this problem. 
A detailed optimization of the parameter choices in the \icss\ scheme is, however, obviously outside the scope of the present work. 

\section{Conclusions}
\label{s:conclusions}

The implementation of the matching between the TMD and collinear factorization theorems, together with a reliable estimation of its uncertainty from power corrections, is one of the compelling milestones for the next generation of phenomenological analyses of $q_\st$-spectra.
This work contributes to such an effort by introducing a new matching scheme: the \emph{inverse-error weighting} (\inew).

From the expected scaling of the power corrections for the TMD and collinear factorization theorems, we build a matched cross section via a weighted average, where the normalized weights are given by the inverse of the (square of the) power corrections.

In the \inew\ scheme, no cancellation of double-counted contributions is needed, since the resummed and fixed-order results are \emph{averaged}, and \emph{not summed}.
This makes the implementation of the cross-section matching in phenomenological analyses faster and more transparent, an important feature in light of the demands of global TMD analyses. 
Moreover, the \inew\ scheme yields compatible results with other mainstream approaches in the literature, such as the improved CSS scheme.

We have illustrated the application of the \inew\ method with the $q_\st$-spectra of $Z$ boson, $W$ boson, $H^0$ boson and Drell-Yan lepton-pair production at the LHC.
However, the \inew\ scheme can be applied in a straightforward manner to any observable where a resummed and a fixed-order factorization theorems need to be matched in order to describe the full spectrum of a given variable, such as the $q_\st$-spectra with polarized beams, event shapes or multi-differential observables.
We leave for the future the study of processes sensitive to (un)polarized TMD fragmentation functions, such as $e^+e^- \to h_1 h_2 X$ and SIDIS, and low-scale processes sensitive to (un)polarized gluon TMDs, such as pseudoscalar quarkonia produced at a future fixed-target experiment at the LHC (AFTER@LHC~\cite{Brodsky:2012vg,Lansberg:2012kf,Boer:2012bt,Kikola:2017hnp}) or even at the LHC~\cite{Aaij:2014bga,Butenschoen:2014dra,Han:2014jya,Zhang:2014ybe,Feng:2015cba,Lansberg:2017ozx}, and the production of a pair of $J/\psi$~\cite{Lansberg:2017dzg}.

{\it \textbf{Acknowledgements.}}
We thank A. Bacchetta, D. Boer, J.C. Collins, P.J. Mulders, J. Qiu, T. Rogers, L. Massacrier, H.-S. Shao, J.X. Wang for useful discussions. 
MGE is supported by the European Research Council (ERC) under the European Union's Horizon 2020 research and innovation program (grant agreement No. 647981, 3DSPIN).
The work of JPL is supported in part by the French CNRS via the LIA FCPPL (Quarkonium4AFTER) and the IN2P3 project TMD@NLO and via the COPIN-IN2P3 agreement.
AS acknowledges support from U.S. Department of Energy contract DE-AC05-06OR23177, under which Jefferson Science Associates, LLC, manages and operates Jefferson Lab. 
TK acknowledges support from the Alexander von Humboldt Foundation and the European Community under the ``Ideas'' program QWORK (contract 320389).

\bibliographystyle{utphys}  
\bibliography{references}

\end{document}